\documentclass{article}

\usepackage{amssymb}
\usepackage{PRIMEarxiv}
\usepackage{amsmath}
\usepackage{amsthm}
\usepackage{amstext}
\usepackage{graphicx}
\usepackage{subcaption}
\usepackage{multirow}
\usepackage[square,numbers]{natbib}

\usepackage[utf8]{inputenc} 
\usepackage[T1]{fontenc}    
\usepackage{hyperref}       
\usepackage{url}            
\usepackage{booktabs}       
\usepackage{amsfonts}       
\usepackage{nicefrac}       
\usepackage{microtype}      
\usepackage{lipsum}
\usepackage{fancyhdr}       
\usepackage{graphicx}       
\graphicspath{{media/}}     

\begin{document}
  
\title{Using deterministic tourist walk as a small-world metric on Watts-Strogatz networks}

\author{
  Joao V. Merenda, Odemir M. Bruno \\
 *Scientific Computing Group \\ Institute of Physics of Sao Carlos \\ University of Sao Paulo \\
  \texttt{joao.merenda@usp.br, bruno@ifsc.usp.br}}
  
\maketitle

\begin{abstract}
The Watts-Strogatz model (WS) has been demonstrated to effectively describe real-world networks due to its ability to reproduce the small-world properties commonly observed in a variety of systems, including social networks, computer networks, biochemical reactions, and neural networks. As the presence of small-world properties is a prevalent characteristic in many real-world networks, the measurement of "small-worldness" has become a crucial metric in the field of network science, leading to the development of various methods for its assessment over the past two decades. In contrast, the deterministic tourist walk (DTW) method has emerged as a prominent technique for texture analysis and network classification. In this paper, we propose the use of a modified version of the DTW method to classify networks into three categories: regular networks, random networks, and small-world networks. Additionally, we construct a small-world metric, denoted by the coefficient $\chi$, from the DTW method. Results indicate that the proposed method demonstrates excellent performance in the task of network classification, achieving over $90\%$ accuracy. Furthermore, the results obtained using the coefficient $\chi$ on real-world networks provide evidence that the proposed method effectively serves as a satisfactory small-world metric.

 \end{abstract}

\keywords{Complex networks \and Watts-Strogatz model \and Small-world networks \and Partially self-avoiding walk \and Pattern Recognition}

\section{Introduction}
\label{sec:introduction}

Network science is an interdisciplinary field of research that encompasses various scientific disciplines such as graph theory, statistics, physics, and computer science. Network science has been applied to investigate a wide range of real-world systems, including social networks \cite{li2019complex, wasserman1994social, axtell2000effects}, the spread of diseases \cite{scabini2021social}, metabolic networks \cite{zhao2006complex, machicao2018topological}, neural networks \cite{stam2007graph}, pattern recognition in texture analysis \cite{chalumeau2007complex, da2006hierarchical}, among others. The use of networks and graph theory to study real-world problems dates back to the 18th century, when Leonhard Euler employed graph theory to solve the problem of the seven bridges of Königsberg. Since then, a plethora of network models have been developed.

In 1959, Paul Erdos and Alfred Renyi introduced an algorithm to generate ensembles of random networks. The Erdos-Renyi (ER) model involves adding edges to the graph with a probability $p$. This can be mathematically represented as $G(N,p)$, where $N$ represents the network size (number of nodes) and $p$ represents the connection probability. Random graphs generated by this model typically exhibit short distances between nodes and a low clustering coefficient. In contrast, regular networks have longer paths between pairs of nodes and a high clustering coefficient. However, many real-world networks, such as neural networks \cite{CHATTERJEE2007145} and fictional character networks \cite{li2019complex}, display the small-world effect, characterized by short diameters, similar to random networks, while also possessing a high clustering coefficient like regular lattices. These networks are referred to as small-world networks \cite{doi:10.1073/pnas.200327197}.

To more accurately model real-world networks, Duncan Watts and Steven Strogatz developed the Watts-Strogatz (WS) model \cite{watts1998collective}. The WS model begins with a circle network, a regular graph with periodic boundary conditions containing $N$ nodes and degree $k$. Then, each edge is rewired with a probability $p$. This process introduces "shortcuts" that connect two nodes previously separated by a long distance, resulting in a decrease in the average path length. If $p = 0$, no edges are rewired and the network remains regular. If $p = 1$, all edges are rewired to random positions, resulting in a random network. For $0 < p < 1$, shortcuts are created, which quickly decrease the average path length, while only slightly affecting the clustering coefficient. As a result, the process generates a hybrid network that closely mimics many real-world networks.

As many real-world networks possess small-world properties, a metric to quantify this effect, also known as small-worldness, has become crucial. There are various ways to define it, one of the most well-known being the $\omega$ (omega) coefficient \cite{telesford2011ubiquity}, which is defined as:

\begin{equation}
\label{eq1}
\omega = \frac{L_{random}}{L} - \frac{C}{C_{lattice}}.
\end{equation}

This metric compares the clustering coefficient ($C$) and the average shortest path length ($L$) of a given network to an equivalent regular network ($C_{lattice}$) and an equivalent random network ($L_{random}$). This quantity varies from $-1$ to $1$, and tends to zero when $L \sim L_{random}$, and $C \sim C_{lattice}$, indicating that the network approximates a small-world network.

In the field of pattern recognition, the Deterministic Tourist Walk (DTW) method has been widely used for texture analysis \cite{backes2006deterministic, gonccalves2012texture, backes2010texture} and network classification \cite{gonccalves2012complex}. The DTW algorithm is an agent-based method, which can be conceptualized as an agent, or "tourist," who visits cities on a map according to deterministic walking rules. In other words, the agent traverses the network by moving from one node to another according to a predefined walking law. The pattern produced by the agent's walk is then used to generate signature vectors for network classification.

In this paper, we propose the use of a modified version of the Deterministic Tourist Walk (DTW) algorithm to investigate the transition between regular and random networks in Watts-Strogatz graphs. Additionally, we utilize the proposed model to construct a new metric for measuring the small-worldness of networks.

\section{Deterministic Tourist Walk}\label{sec:DTW}

The Deterministic Tourist Walk (DTW) is an agent-based method in which an agent, or "tourist," visits nodes according to a predefined walking rule. In the traditional DTW algorithm, which has been applied to texture analysis \cite{backes2006deterministic}, the agent can move according to two rules: the rule of minimum, in which the agent moves towards the node that minimizes a specific measure (e.g. degree difference between two nodes); and the rule of maximum, in which the agent aims to maximize this measure.

The trajectory of the agent consists of two parts: the transient, which has a length of $t$, and the attractor, which has a length of $a$, and occurs when the agent gets stuck in a loop. The total trajectory length is the sum of these two parts. The walker has a memory size ($\mu$) that stores the last $\mu$ nodes it has visited. The agent is not allowed to return to a node stored in its memory, resulting in a partially self-avoiding walk. The minimum attractor length is $ a = \mu + 1$.



In the proposed method, the agent, or "walker," is launched from every node in the network, resulting in an ensemble of $N$ trajectories, each with a unique value of transient and attractor. Thus, a 2-dimensional histogram illustrating all possible combinations of transient and attractor can be calculated using the following equation:

\begin{equation}
\label{eq2}
h(\ell) = \sum_{b = 1}^{\ell}S(b, \ell - b)
\end{equation}

where $\ell$ is the trajectory length and $S(b, \ell - b)$ is a counter function that counts the number of trajectories with transient $t = b$, and attractor size $a = \ell -b$. The $S(t,a)$ function is defined as:

\begin{equation}
    \label{eq3}
    S(t, a) = \frac{1}{N}\sum_{i = 1}^N
    \begin{cases}
        1, & \text{if $t_i = t$, and $a_i = a$} \\
        0, & \text{otherwise}
    \end{cases}
\end{equation}

In other words, this function calculates the proportion of trajectories with a given transient and attractor size out of the total number of trajectories.

%
%
%

From this histogram, a signature vector $\phi (\mu,\ r)$, as shown in Eq. \ref{eq4}, is constructed for agents with memory size $\mu$ and walking rule $r$. Traditionally, the first element of the signature vector is the histogram for trajectories with length $\mu + 1$.

\begin{equation}
\label{eq4}
\phi (\mu,\ r) = [h(\ell = \mu + 1),\ h(\ell = \mu + 2),\ h(\ell = \mu + 3),\ \cdots ,\ h(\ell = N)]
\end{equation}

Several variations of $\phi (\mu,\ r)$ have been proposed. In \cite{gonccalves2012texture}, Gonçalves \textit{et al.} proposed concatenating the $\phi (\mu,\ r)$ vector for multiple memory values to increase the efficiency of the DTW algorithm in classification tasks. This new vector ($\psi$) is defined by:

\begin{equation}
\label{eq5}
\psi (\vec{\mu},\ r) = \psi ([\mu_1,\ \mu_2,\ \cdots ,\ \mu_m],\ r) = [\phi (\mu_1,\ r),\ \phi (\mu_2,\ r),\ \cdots ,\ \phi (\mu_m,\ r)]
\end{equation}

\section{Modified tourist walk}
\label{sec:proposed}
In contrast to the traditional Deterministic Tourist Walk (DTW) method, which is a general approach for any network model, the DTW algorithm presented in this paper includes two modifications to tailor the method for the Watts-Strogatz model. The first modification is in the walking rules, and the second modification is in the construction of the signature vector.

\subsection{Walking rules}
%
The first rule is a necessary condition for the agent's movement. The agent moves only if there is a node in its vicinity that has a degree different from the degree of its current node. Mathematically, let $k_j$ be the degree of the $j-th$ neighbor, and $k_i$ be the degree of the current node, then:

\begin{equation}
\label{eq6}
W_{candidates, i} = { \forall j \in W_i : |k_i - k_j| \neq 0 }
\end{equation}

where $W_i$ is the set of neighboring nodes of node $i$, and $W_{candidates, i}$ is the set of "candidate" neighboring nodes that the agent can choose to visit. The main consequence of this first rule is that if the network is completely regular, \textit{i.e.}, $W_{candidates, i} = {\emptyset }$ for all nodes in the network, then the agent does not move. In this case, the algorithm returns a transient value ($t$) of zero and an attractor value ($a$) of zero. If the first rule is satisfied, \textit{i.e.}, $W_{candidates, i} \neq {\emptyset }$, the agent moves towards the node that minimizes the difference in the clustering coefficient, as shown in Eq. \ref{eq7}. This is the second walking rule.

\begin{equation}
    \label{eq7}
    rule\ =\   min([\ |c_i - c_j|\ ]),\ \forall j \in W_{candidates, i}
\end{equation}
where $c_i$ is the clustering coefficient of the current node, and $c_j$ is the clustering coefficient of the $j-th$ candidate node.

There are three stop conditions for the agent's walk: The first one occurs when the agent reaches a locally regular part of the network, and in this case, the algorithm returns a transient value $t = \tau$, and an attractor value $a = 0$, or $t = 0$, and $a = 0$ if the network is entirely regular. The second stop condition occurs when the agent reaches an attractor, and the algorithm returns a transient value $t = \tau$, and an attractor value $a = \alpha$. The last stop condition occurs if the agent does not find an attractor and exceeds the maximum number of iterations allowed (\textit{e.g.}, the network size). In this case, the algorithm returns a transient value $t = N$, and an attractor value $a = 0$.

\subsection{The signature vector}
In order to adapt the method to these new rules, the term $h(\ell = 0)$ was added to the vector defined by Eq. \ref{eq4}, so the signature vector is now given by
\begin{equation}
    \label{eq8}
    \phi (\mu ) = [h(\ell = 0),\ h(\ell = \mu + 1),\  h(\ell = \mu + 2),\ \cdots\ ,h(\ell = N)]
\end{equation}
where $h(\ell = 0)$ is a "regularity meter" which is maximum ($h(\ell = 0) = 1$) when the network is completely regular and tends to zero in random networks. The expanded signature vector defined by Eq. \ref{eq5} remains the same.

\subsection{A small-world metric from the DTW algorithm}
Let be $C$ the global clustering coefficient of the network $G(V, E)$, and $\bar{\ell}$, the average trajectories length of the tourist walk upon $G$. Thus, the tourist small-worldness coefficient  $\chi$ is defined as 

\begin{equation}
    \label{eq9}
    \chi = C.\frac{\bar{\ell}_G}{\bar{\ell}_{Rand}}
\end{equation}
where $\bar{\ell}_{Rand}$ is the average length of the trajectories yielded by the tourist in an equivalent random network.

\section{Experimental}
\label{sec:experimental}
\subsection{Network datasets}

In order to test the proposed method was used three datasets. 

\begin{enumerate}
    \item Synthetic-dataset: This synthetic dataset comprises of three network models: i) Random networks, generated using the Erdos-Renyi model, with a probability of connection equal to $p = \langle k \rangle / (N-1)$; ii) Regular networks, and iii) small-world networks, generated using the Watts-Strogatz model, with $0.01 \leq p \leq 0.1$. The experiments were conducted using four network size values: $N =$ 500, 1000, 1500, 2000, and seven average degree values: $\langle k \rangle =$ 4, 6, 8, 10, 12, 14, 16. Each of these 28 combinations of $N$ and $\langle k \rangle$ has 100 networks, resulting in 2800 networks for each model and a total of 8400 networks.
    
    \item Noisy-dataset: To evaluate the robustness of the proposed method, noise was introduced to the synthetic-dataset at various rates. Specifically, the experiments were conducted using three noise levels: $10\%$, $20\%$, and $30\%$. The noise was applied by randomly removing or adding edges within the network.
    
    \item Real-world networks: This dataset is composed of networks extracted from the KONECT project \cite{kunegis2017konect}.
    
\end{enumerate}

\subsection{Classification method}
The Linear discriminant analysis (LDA) classifier was employed in the experiment to classify the network datasets. The Leave-one-out cross-validation (LOOCV) approach was used to evaluate the prediction capacity of the model. In the LOOCV method, the dataset is divided into $n-1$ equal parts, with one data point left out each time. The subset is then classified, and the model returns the average accuracy and the standard deviation.

\subsection{Comparison method}
A vector $\Lambda$ containing structural measures of the network was constructed to compare the performance of the proposed method. Four network measures, along with their respective means and standard deviations, were utilized: degree, clustering coefficient, shortest path length, and assortativity. As a result, the vector $\Lambda$ comprises eight elements.

\subsection{Runtime comparison}
In the runtime comparison, all experiments were performed using the same computational setup. Intel (R) i7-7820X 32-cores 3.2 GHz, 96 GB RAM, and an Nvidia (R) GTX 1080Ti 11 GB graphic processor unit.

\section{Results and discusion}
\label{sec:results}

\begin{figure}[h!]
\centering
\includegraphics[scale=0.9]{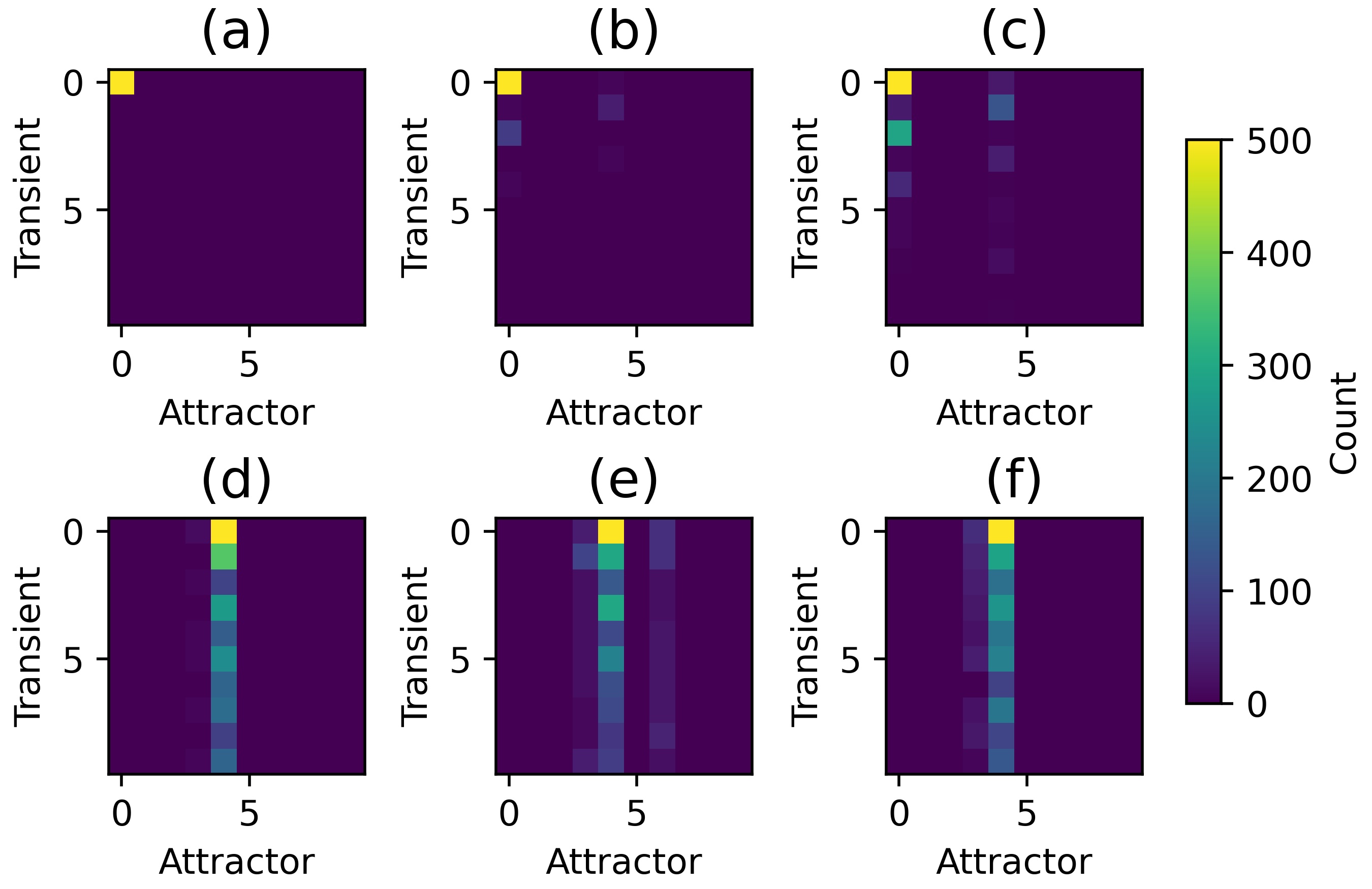}
\caption{Two-dimensional histogram of the trajectories yielded by the tourist upon a network with $500$ nodes and average degree equals $10$. The tourist has memory size $\mu = 2$. The outcomes were obtained for some rewiring probabilities $p$: (a) $p = 0.0001$. (b) $p = 0.001$. (c) $p = 0.01$. (d) $p = 0.05$. (e) $p = 0.1$. (f) $p = 1.0$.
}
\label{fig1}
\end{figure}

The proposed model results in non-zero trajectories only for networks that do not possess regular properties. Thus, the two-dimensional histogram (transient X attractor) should exhibit a maximum density at the point $(0,0)$ for regular networks. For random networks, the tourist has a high degree of freedom, and can present trajectories of different sizes in various combinations of transient and attractor lengths. For a rewiring probability of $p = 1.0 \times 10^{-4}$ (as shown in Figure \ref{fig1}(a)), the network is almost regular and exhibits a high density at the point $(0,0)$. Conversely, for $p = 1.0$ (as shown in Figure \ref{fig1}(f)), there is no trajectory of length zero. In the interval $0.01 \leq p \leq 0.1$ (as shown in Figures \ref{fig1}(c) to \ref{fig1}(e)), a transition occurs, where the network ceases to possess regular properties and becomes increasingly random. However, within this interval, the networks still possess properties of both models, as they are small-world networks.

\subsection{Classification}
\begin{figure}[!htb]
\centering
\includegraphics[scale=0.8]{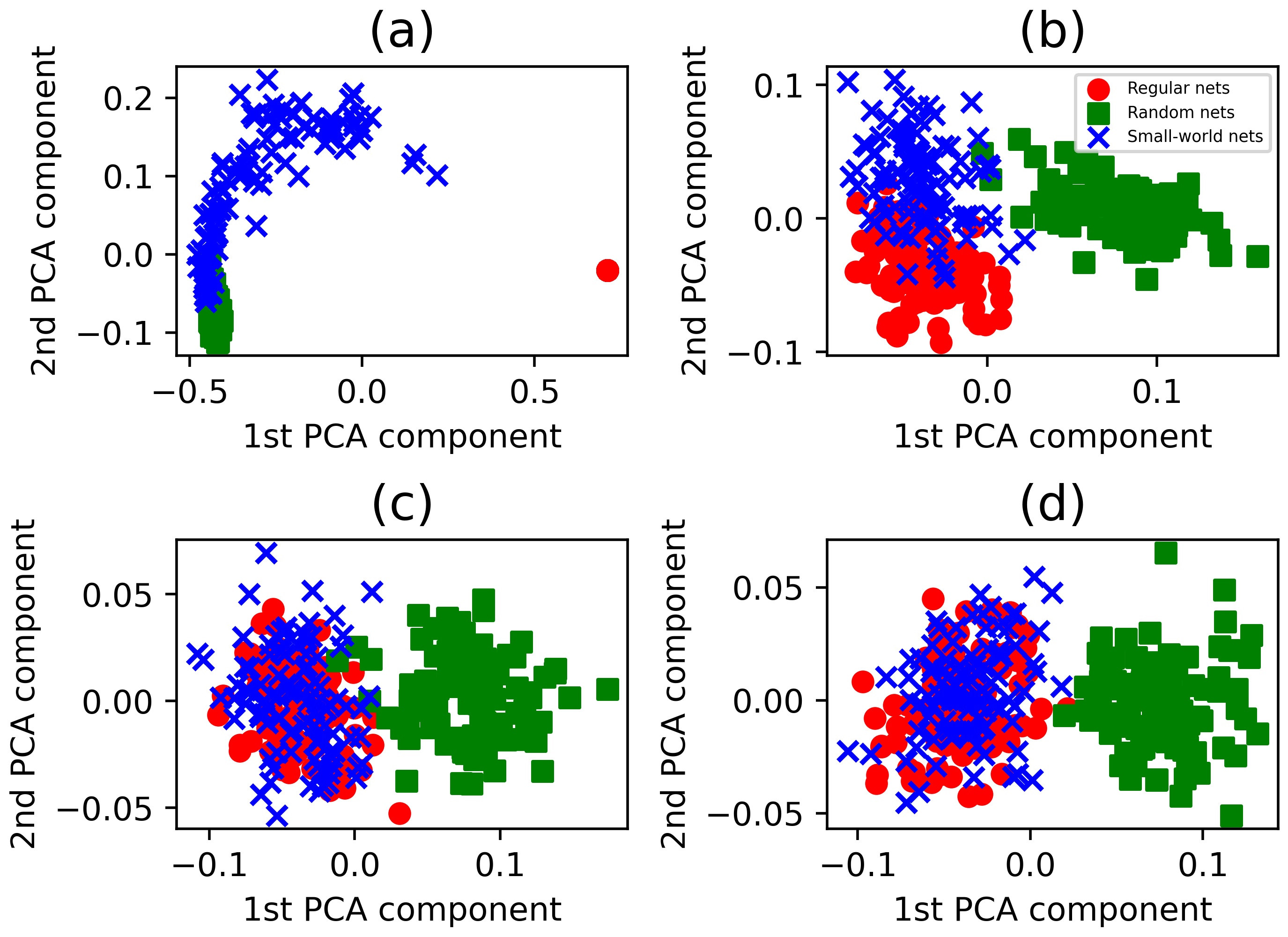}
\caption{The classification of a sample of the synthetic-dataset by the PCA method. The sample contains $300$ networks divided equally into three models: regular networks (in red), random networks (in green), and small-world networks (in blue). (a) Free-noise networks. (b) Networks with $10\%$ of noise. (c) Networks with $20\%$ of noise. (d) Networks with $30\%$ of noise.
}
\label{fig2}
\end{figure}

Considering the use of only one memory size $\mu$, Table \ref{table1} illustrates that the proposed method demonstrates satisfactory performance in comparison to the structural measures of the network. Utilizing a memory size of $\mu = 1$ yielded the best result, achieving an accuracy of $97\%$ for noise-free networks, and $93\%, 87\%, 84\%$ for networks with noise rates of $\rho = 10\%, 20\%, 30\%$ respectively. The performance of the DTW method decreases as the memory size increases. The classification of a sample of the dataset containing 300 networks is illustrated in Fig. \ref{fig2}. The addition of noise transforms regular networks into small-world networks and some small-world networks into random networks (Fig \ref{fig2}(b)(c)(d)). As such, it is expected that the method's performance will decrease as noise is added to the network. However, as shown in Table \ref{table1}, this decrease is not significant and the DTW method maintains an accuracy of over $80\%$ even for networks with $30\%$ noise.

\begin{center}
\begin{table}[h!]
\begin{tabular}{ c c c c c}
\hline
 $\phi (\mu )$& synthetic-dataset &  \multicolumn{3}{c}{Noisy-dataset}   \\ 
 \cline{3-5}
  & & $\rho=10\%$ & $\rho=20\%$ & $\rho=30\%$\\
 \hline
 $\phi (1)$ & $97\ (\pm\ 1)$ & $93\ (\pm\ 3)$ & $87\ (\pm\ 4)$ & $84\ (\pm\ 4)$ \\  
 $\phi (2)$ & $97\ (\pm\ 1)$ & $92\ (\pm\ 2)$ & $86\ (\pm\ 3)$ & $82\ (\pm\ 3)$ \\ 
 $\phi (3)$ & $97\ (\pm\ 1)$ & $91\ (\pm\ 2)$ & $87\ (\pm\ 3)$ & $83\ (\pm\ 3)$ \\ 
 $\phi (4)$ & $95\ (\pm\ 1)$ & $86\ (\pm\ 3)$ & $84\ (\pm\ 3)$ & $81\ (\pm\ 3)$ \\ 
 $\phi (5)$ & $94\ (\pm\ 1)$ & $85\ (\pm\ 3)$ & $82\ (\pm\ 4)$ & $79\ (\pm\ 4)$ \\ 
 Structural measures & $98\ (\pm\ 1)$ & $96\ (\pm\ 2)$ & $93\ (\pm\ 2)$ & $91\ (\pm\ 3)$ \\
 \hline
 
\end{tabular}
\caption{Accuracy ($\%$) and standard deviation for the single-memory tourist approach.}
\label{table1}

\end{table}
\end{center}

\begin{center}
\begin{table}[!h]
\begin{tabular}{ l c c c c}
\hline
 $\psi (\vec{\mu})$& synthetic-dataset &  \multicolumn{3}{c}{Noisy-dataset}   \\ 
 \cline{3-5}
  & & $\rho=10\%$ & $\rho=20\%$ & $\rho=30\%$\\
 \hline
 $\psi([1,\ 2])$ & $99\ (\pm\ 1)$ & $96\ (\pm\ 2)$ & $94\ (\pm\ 2)$ & $90\ (\pm\ 2)$ \\  
 $\psi([1,\ 2,\ 3])$ & $99\ (\pm\ 1)$ & $95\ (\pm\ 2)$ & $97\ (\pm\ 2)$ & $92\ (\pm\ 3)$ \\ 
 $\psi([1,\ 2,\ 3,\ 4])$ & $99\ (\pm\ 1)$ & $99\ (\pm\ 1)$ & $99\ (\pm\ 1)$ & $97\ (\pm\ 1)$ \\ 
 $\psi([1,\ 2,\ 3,\ 4,\ 5])$ & $100\ (\pm\ 0)$ & $100\ (\pm\ 0)$ & $99\ (\pm\ 1)$ & $99\ (\pm\ 1)$ \\ 

 \hline
 
\end{tabular}
\caption{Accuracy ($\%$) and standard deviation for the multiple-memories tourist approach.}
\label{table2}

\end{table}
\end{center}

When combining multiple memory sizes, the performance of the proposed method improved, as demonstrated in Table \ref{table2}. Notably, there is a significant gain in accuracy for the noisy dataset. Small memory sizes allow the tourist to extract microscale information from the network, while larger memory values enable the tourist to scan a larger region of the network. By combining several memory values, the deterministic tourist walk method is able to perform a multiscale analysis of the network, thereby extracting more meaningful information about it.

\subsection{Small-worldness measures}
\begin{figure}[!htbp]
\centering
\includegraphics[scale=1.0]{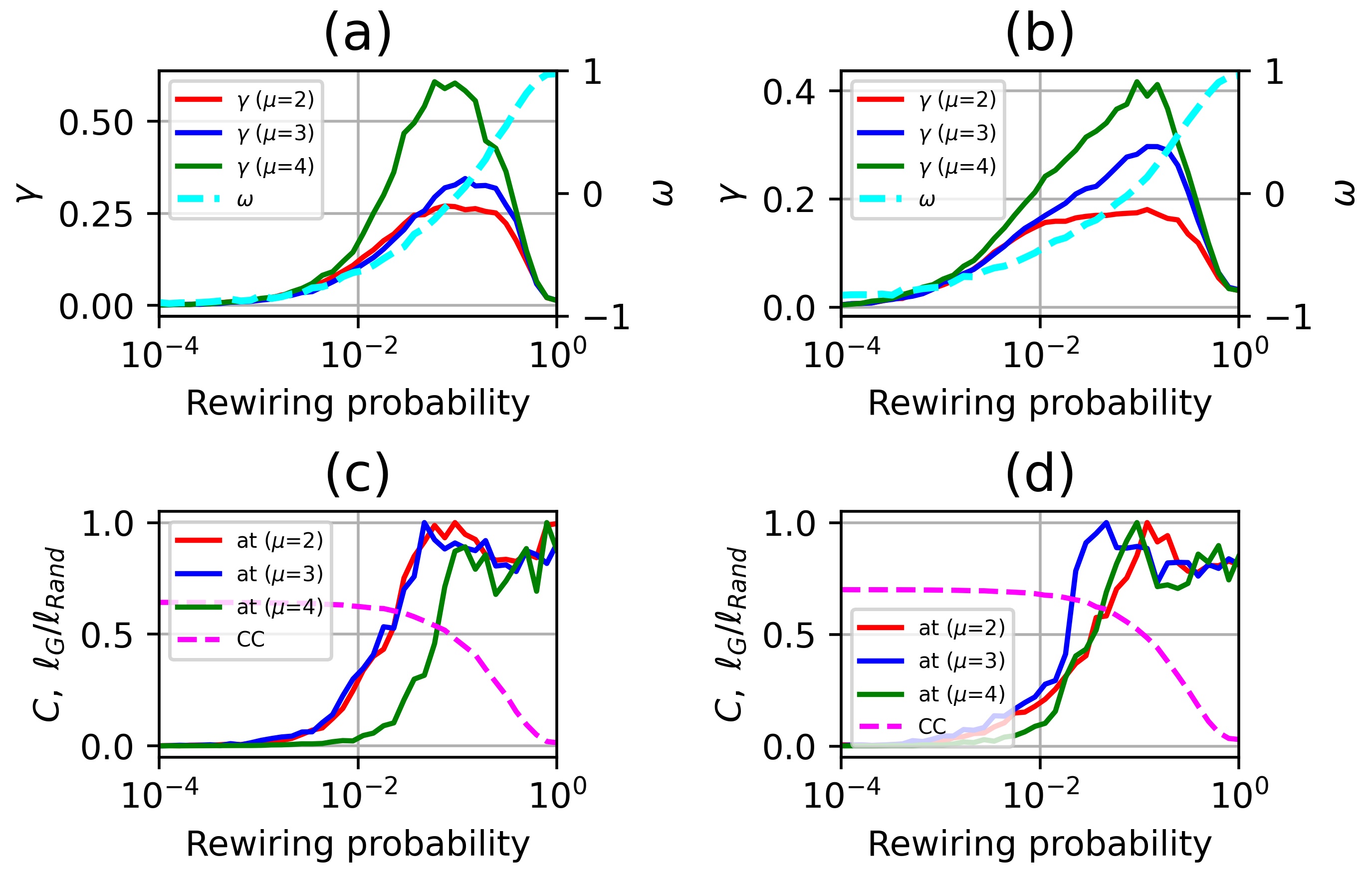}
\caption{
The $\chi$ coefficient for Watts-Strogatz networks with rewiring probability ranging from $10^{-4}$ to 1.0. The red, blue, and green lines represent the DTW method with memory $\mu = 1$, $\mu = 2$, and $\mu = 3$, respectively. The dashed line in magenta represents the clustering coefficient, and the dashed line in cyan represents the $\omega$ coefficient. (a) compares the $\chi$ coefficient (left axis), and the $\omega$ coefficient (right axis) for networks with 500 nodes and average degree $\langle k\rangle = 8$. (b) compares the $\chi$ coefficient (left axis), and the $\omega$ coefficient (right axis) for networks with 500 nodes and average degree $\langle k\rangle = 16$. (c) compares the clustering coefficient and the average length of the DTW trajectories for networks with 500 nodes and average degree $\langle k\rangle = 8$. (d) compares the clustering coefficient and the average length of the DTW trajectories for networks with 500 nodes and average degree $\langle k\rangle = 16$.
}
\label{fig3}
\end{figure}

The coefficient $\chi$, defined by Eq. \ref{eq9}, comprises both a structural property of the network, the global clustering coefficient, and a property of the proposed tourist walk method, the trajectory length. As the network becomes more regular or small-world, the global clustering coefficient remains high, while it decreases rapidly as the network becomes more random. In contrast, the average length of the DTW trajectories is zero for regular networks (as the tourist cannot move) and tends to have high values for small-world and random networks (as the tourist has more degree of freedom). Therefore, a peak in the $\chi$ curve is expected when both of these measures have high values. Figure \ref{fig3}(c) and (d) show the evolution of these two measures with the rewiring probability. Figure \ref{fig3}(a) and (b) show the coefficient $\chi$ (solid lines) and the coefficient $\omega$ (dashed lines) within the same range of probability. The coefficient $\chi$ presents a peak close to the point where the $\omega$ coefficient is zero, indicating that the $\chi$ coefficient tends to be the maximum for small-world networks.

As shown in Table \ref{table4}, the $\chi$ values for seven real-world networks are presented. According to the $\omega$ coefficient, the Word-adjacency network is more random than the others, thus it is expected that its $\chi$ value is low. This network presents the smallest $\chi$ coefficients in the table. The \textit{C. elegans} (metabolic) network is more small-world than the \textit{C. elegans} (neural) network. As a result, the first one has a higher $\chi$ coefficient than the second one for all memory values.

\begin{center}
\begin{table}[h]
\begin{tabular}{ l c c c c c c}
\hline
Network  & C & $\omega$ & $\gamma_{\mu = 1}$ & $\gamma_{\mu = 2}$ & $\gamma_{\mu = 3}$ & $\gamma_{\mu = 4}$ \\
\hline
Jazz & 0.61 & 0.009 & 0.64 & 0.59 & 0.44 & 0.38  \\
Karate club & 0.57 & 0.05 & 0.59 & 0.47 & 0.41 & 0.45  \\
\textit{C. elegans} (metabolic) & 0.64 & 0.18 & 0.56 & 0.34 & 0.27 & 0.32 \\
Football (Girvan-Newman) & 0.40 & 0.32 & 0.39 & 0.27 & 0.28 & 0.29 \\
\textit{C. elegans} (neural) & 0.29 & 0.59 & 0.31 & 0.15 & 0.23 & 0.22 \\
Word adjacency & 0.17 & 0.73 & 0.16 & 0.073 & 0.08 & 0.13 \\
 \hline
 
\end{tabular}
\caption{Real-world networks (extracted from \cite{kunegis2017konect}) with their respective clustering coefficient ($C$), $\omega$ coefficient, and $\chi$ coefficient for memories values: 1, 2, 3, and 4.}
\label{table4}

\end{table}
\end{center}

\begin{figure}[!htbp]
\centering
\includegraphics[scale=0.8]{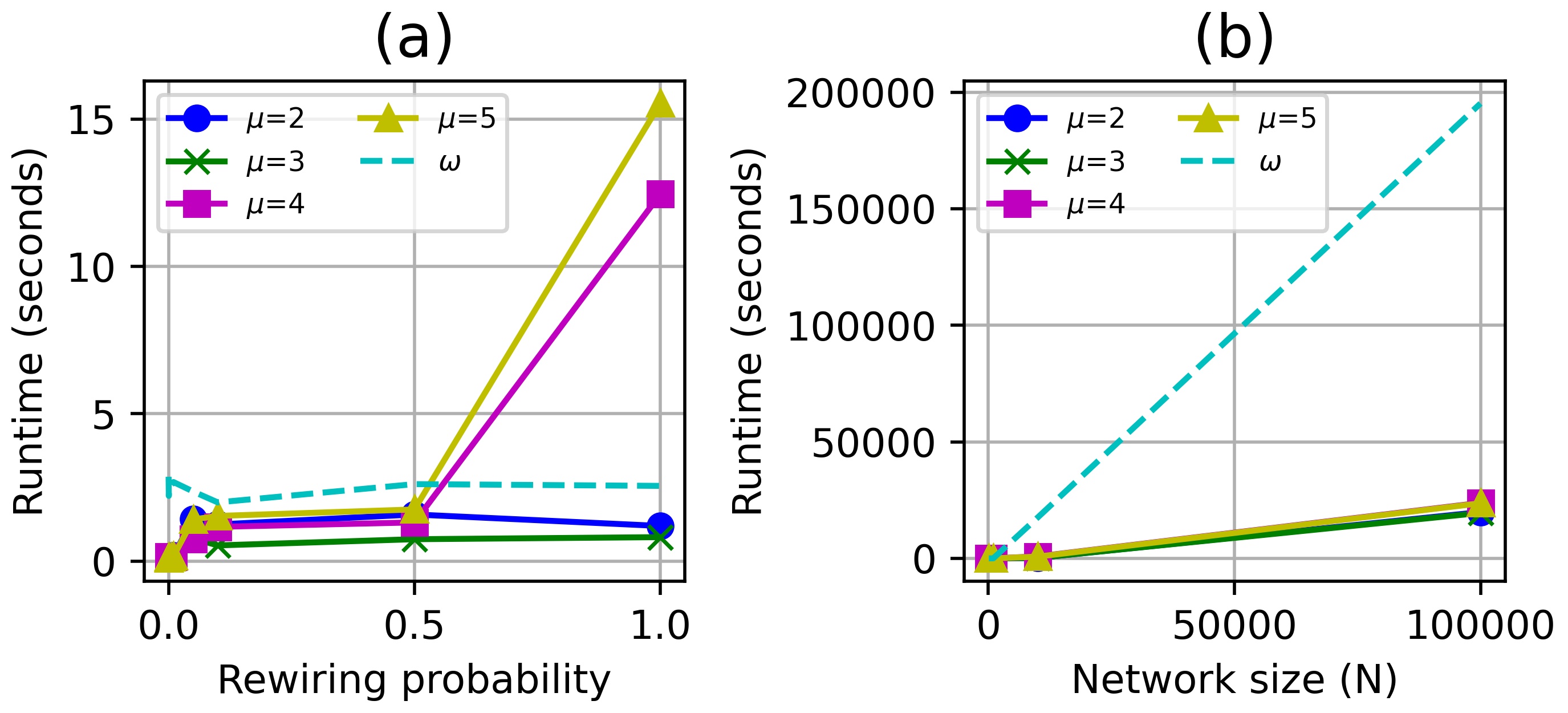}
\caption{(a) Runtime comparison between $\chi$ and $\omega$ coefficients for networks with 500 nodes and an average degree equal to 10. The rewiring probability assumes the following values:
$p = 0.0001,\ 0.0005,\ 0.001,\ 0.005,\ 0.01,\ 0.05,\ 0.1,\ 0.5,\ 1.0$. (b) Runtime comparison for $\chi$ and $\omega$, varying the network size: $N = 10^2,\ 10^3,\ 10^4,\ 10^5$. The rewiring probability is $p = 0.05$.
}
\label{fig4}
\end{figure}

In Figure \ref{fig4}, the comparison of the runtime between the $\chi$ coefficient and the $\omega$ coefficient is shown. With the network size as a constant, and the rewiring probability as a variable, in this first case, the two methods have similar speeds, as illustrated in Figure \ref{fig4}(a), except for DTW with memory $\mu = 4$ and $\mu = 5$. However, the significant advantage of the proposed method is observed for large networks, as shown in Figure \ref{fig4}(b). As the size of the network increases, the calculation of the $\chi$ coefficient proves to be faster than the $\omega$ coefficient. All runtime comparisons were performed using the same computational setup.
  
\section{Conclusion}\label{sec:conclusion}

In this paper, a new approach for extracting meaningful information from Watts-Strogatz networks using a partially self-avoiding walk, the deterministic tourist walk, is presented. The novel version of the DTW method introduced here achieved good performance in the classification of networks and highlighted the transition from regular networks to random networks in the WS model. Furthermore, the proposed method also exhibited good results when used as a metric for small-worldness. The small-world effect is prevalent in many real-world networks, making the identification of a metric to quantify this effect crucial. Unlike other metrics that consider only structural properties of the network, the proposed method reached good results by combining a structural property (clustering coefficient) with properties of an automaton (trajectory length). Further studies should be conducted in the future to enhance this method.

\section*{Acknowledgements}
J.V. Merenda gratefully acknowledges the financial support grant \\$\#88887.601525/2021-00$, Coordination for the Improvement of Higher Education Personnel (CAPES). O. M. Bruno acknowledges support from CNPq (Grant \#307897/2018-4) and FAPESP (grants \#2018/22214-6 and \#2021/08325-2). The authors are also grateful to the NVIDIA GPU Grant Program.


\end{document}